\documentstyle[12pt,draft]{article}

 \newcommand \be {\begin{equation}}
\newcommand \bea {\begin{eqnarray} \nonumber }
\newcommand \ee {\end{equation}}
\newcommand \eea {\end{eqnarray}}

\newcommand \la {\lambda}

 \newcommand \al {\alpha}
 \newcommand \NE {\not=}
 \newcommand \N {{\cal N}}
\newcommand \R {{\cal R}}
\newcommand \LL{{\cal L}}
\newcommand \ba {\overline}
\newcommand \lan {\langle}
\newcommand \ran {\rangle}
\begin{document}

\title{$D$-dimensional Arrays of Josephson Junctions, Spin Glasses and
$q$-deformed Harmonic
Oscillators}
\author{  Giorgio Parisi \\
Dipartimento di Fisica, Universit\`a {\sl La  Sapienza}\\
INFN Sezione di Roma I \\
Piazzale Aldo Moro, Rome 00185}
\maketitle

\begin{abstract}
We study the statistical mechanics of a $D$-dimensional
array of Josephson junctions
in presence of a magnetic field. In the high temperature region the
thermodynamical properties can be
computed in the limit $D \to \infty$, where the problem is simplified;
this limit is taken in the framework of
the mean field approximation. Close to the transition point the system
behaves very similar to a particular form of spin
glasses, i.e. to gauge glasses. We have noticed that in this limit the
evaluation of the coefficients  of the high temperature
expansion  may be mapped onto the computation of some matrix
elements
for the $q$-deformed
harmonic oscillator.
 \end{abstract}
\vfill
\vfill
\newpage

\section {Introduction}

In this paper we are interested to study the statistical mechanics of
arrays of Josephson junctions
in $D$-dimensions in the limit where $D \to \infty$. We will construct
here the solution of the mean
field theory in  the high temperature phase. We postpone to a later stage
the computation of the
corrections to  the mean field approximation and the study of the low
temperature phase. The model
has been studied in two dimensions, especially in the low temperature
region \cite{HAL,BEL}, but no
results are known in very high dimensions.

The model we  consider is described by the Hamiltonian:
\be
H= - c(D) \sum_{j,k} {\ba \phi_j } U_{j,k} \phi_k + h.c.\ .
\ee
Here $c(D)$ is a normalisation constant, which will be useful later to
rescale the Hamiltonian in
order to obtain a non trivial limit when $D$ goes to infinity. The spins
$\phi_j$ are defined on a
$D$-dimensional hypercubic lattice.

We can consider three possibilities:
\begin{itemize}
\item  The  spins $\phi_j$ are constrained to be of modulus one.
\item The spins $\phi_j$ have  modulus one in the average at $\beta=0$:
in this limit
they have a Gaussian distribution.
\item The spins satisfy the constraint $\sum_i|\phi_i|^2 =N$. This is the
spherical model which is intermediate among the two previous models.
\end{itemize}

In the limit where the dimension $D$ goes to infinity the
properties of the  first model and of the third model can be
obtained from that of the Gaussian model. We will concentrate our
attention on the Gaussian case.

 The couplings $U$
are non zero only for nearest neighbour sites. They are complex
numbers of modulus one and they
satisfy the relation
\be
U_{k,j} = \ba{ U_{j,k} }.\label {sym}
\ee
In other words  the couplings $U$ are the links variables of an
$U(1)$ lattice gauge field.

We will select the couplings $U$ to give a constant
magnetic field.
Many different orientations of the magnetic field can be chosen. For
simplicity we restrict our computation to
the  case where the flux through each elementary plaquette is given by
$B$ (or $-B$), independently
from the plane to which the plaquette belongs. This
corresponds to
constant uniform frustration on all the plaquettes. In the extreme case
($B=\pi$) we obtain a fully
frustrated  model, while for $B=0$ we recover the ferromagnetic case.
Random point dependent $B$ values
correspond to  a particular form of spin
glasses, i.e. to gauge glasses \cite{GG4}-\cite{PARISI}.

 More precisely we set
 \be
B_{\al,\beta} = S_{\al,\beta} B ,
\ee
where $S_{\al,\beta}$ may take the values 1 or $-1$, $B_{\al,\beta}$ is the
antisymmetric tensor corresponding to the  magnetic field, which in the
continuum limit is given by $\partial_\al A_\beta - \partial_\beta A_\al$.
The ordered product of the
four links of a plaquette in the $\al,\beta$ plane is equal to
$\exp(i\ B_{\al,\beta})$.

We must now specify $S_{\al,\beta}$, i.e. the  sign of $B_{\al,\beta}$. A
possible choice would be to take
 \be
S_{\al,\beta} = 1 \ \ for \  \ \al>\beta,
\ee
which  implies $B_{\al,\beta} = B$ for $\al>\beta$.

In two and in three dimensions this choice is equivalent to any other
possible choice of the sign. In three dimension the magnetic field is a
vector and all the vectors corresponding to different choices of the sign
 may be obtained one from the other with a rotation. The choice of $S$ does
not influence the thermodynamics.

 In more than three dimensions different
choices of the matrix $S$ are not equivalent \footnote {I am grateful to E.
Marinari and F. Ritort for crucial discussions on this problem.} and we must
select one among all the possible ones. In this note we consider the case in
which the matrix $S$ is a generic one, i.e. the signs of $B$ are randomly
chosen.  The system is translation invariant and the randomness appears in
only in the relative orientation of the magnetic field with the crystal axis.


In the two dimensional case we recover the usual description for an $XY$
system (or equivalently an array
of Josephson junctions) in constant magnetic field.

The aim of this note is to compute the statistical properties of this model
in the mean field
approximation in the high temperature region. The first difficulty we
face
consists in finding the spectral
properties of the lattice discretised Laplacian in presence of a magnetic
field. The lattice
Laplacian is defined as \be
(\Delta f)_j=  \sum_k U_{j,k} f_k.
\ee

The spectral properties of the lattice Laplacian in two dimension have
been carefully studied.  They depend on the arithmetic
properties of the  $B/\pi$, i.e. different results are
obtained for rational and irrational $B/\pi$ \cite{BEL}.

 The study of
the lattice Laplacian in higher
dimensions is much less developed. In any dimension the explicit
construction of the field $U$ show
that for rational $B/\pi$, of the form $B=2 \pi r/s$, with both $r$ and
$s$
integers, there is a gauge in
which the $U$ couplings are periodic functions of the position, with
period $s$. In this case the
spectrum of the Laplacian has the typical band form, the edges of the
bands being related to the
eigenvalues of a $s^D \times \  s^D$ matrix. When both $s$ and $D$
are
large, a direct study of the
eigenvalues is rather complex.

 We will study this problem in the limit of an infinite number of
dimensions . We cannot
solve
it in a completely  satisfactory way, but we can put forward some
educated guesses. We will find some
unexpected relations with the properties of the $q$-deformed harmonic
oscillator. At the end  the
behaviour of the model will come out very similar to that of spin
glasses.
 The reader should notice that it is not clear how much of our
results survives in a large, but
finite, dimensions and that the properties of the model in high dimensions
may be quite different from
that of the  two dimensional model.

In section II we present some general considerations. In the next section
we show some general
properties of the high temperature expansion in the limit $D\to \infty$.
We consider in detail the
ferromagnetic case, the spin glass case and the constant frustration
model. In section IV we show the
relation among the high temperature expansion for the constant
frustration model  in infinite
dimension and the $q$-deformed harmonic oscillator. In the next
section we study the behaviour of our
model near the critical point and we find that it is very similar to that of
spin glasses.
In section VI we briefly discuss the problems related to the exchange of
limits ($\beta \to \beta_c$ and $D \to \infty$). Finally (in the last
section) we present our
conclusions and express our points of view on the  open problems. In
the appendix we will describe some interesting features of the
$q$-deformed harmonic oscillator, which shows an anomalous
behaviour for
$q=\exp(2\pi \ i \theta)$, when $\theta$ is rational.

  \section {General Considerations}

There are two extreme cases for the $U$ which are very well studied
for the
Hamiltonian (1):
\begin {itemize}
\item
We set
\be
U_{j,k}=1.
\ee
 In this way we obtain the usual ferromagnetic $XY$ model. There is a
ferromagnetic  transition at
$\beta=1$  in the limit $D \to \infty$, if we set $c(D)= {1 \over 2D}$,
i.e. $c(D)$ has to be equal to the
inverse of the  coordination number of the hypercubic lattice.
\item
We set
\be
U_{j,k}= \exp (i r _{j,k}),
\ee
where $r$ are random numbers belonging to the interval $0-2 \pi$,
such that
the  symmetry condition eq. (\ref{sym})
is satisfied.

In this way be obtain an spin glass model of the $XY$ type, which is called a
gauge glass.  The  transition temperature is  $\beta=1$
in  the limit $D \to \infty$, if we set $c(D)=  (2D)^{-1/2} $,
i.e. \cite{GG4,GG3,GG5} $c(D)$
is equal to the inverse of
the square root of the coordination number.
\end {itemize}

The model we  study is intermediate among the previous two problems.
In order to define
it properly, it is convenient to introduce the so called Wilson loop. Let
us
consider a closed oriented circuit
($C$) on the lattice,  which goes from the point $j$ to the same point
$j$ and let us define $W(C)$ as
the  product of the $U$'s along the circuit. The Wilson loop $W(C)$ is
a gauge invariant. The
knowledge of $W(C)$ for any $C$ gives all gauge invariant
informations concerning the gauge field.

In the continuum limit we have
\be
W(C)=\exp(i\int_C dx^\mu A_\mu(x)) =\exp (i\Phi(C)),
\ee
where  $\Phi(C)$ is the magnetic flux entangled within $C$.

In 2 dimensions in presence of a constant magnetic field the Wilson loop
is given by
\be
W(C)= \exp(i B S(C)),
\ee
where $S(C)$ is the signed area of the loop $C$.

In $D$ dimensions there are  $D(D-1)/2$  planes oriented in the
directions of the lattice. The
choice of the magnetic field we study here is

\bea
W(C)= \exp(i \Phi(C)),  \\
\Phi(C)= \sum_{\nu,\mu\ = \nu<\mu}S_{\nu,\mu}(C)B_{\nu,\mu}\ ,\label{SIGNED}
\eea
where the indices $\nu$ and $\mu$ denote one of the $D$ possible
different directions and
$S_{\nu,\mu}$ is the signed area of the projection of the curve $C$ on
the $\nu,\mu$ plane.

As a consequence of gauge invariance there are infinite many choices of
the $U$ which correspond
to these Wilson loops. All these choice are physically equivalent. In two
dimensions we could set
\bea
U_1(j)=1, \ \ \  U_2(j)= \exp (i B j_1),
\eea
 where $j_\nu$ is the $\nu^{\rm th}$ component of the vector $j$ and
we have introduced the
short-handed notation
\be
U_\nu(j)=U(j,j+n_\nu),
\ee
$n_\nu$ being the unit vector in the $\nu$ direction.

This construction can be generalised to the $D$-dimensional case. For
example in $4$ dimensions
one obtains
\bea
  U_1(j)&=&1, \ \ \  U_2(j)= \exp (i B_{2,1} j_1),\\
  \ \ \  U_3(j)&=& \exp (i ( B_{3,1}j_1+ B_{3,2}j_2)), \ \ \
  U_4(j)= \exp (i ( B_{4,1}j_1+ B_{4,2}j_2+ B_{4,3}j_3)).
\eea

Our main task will be the study of the associated Gaussian model, where
the Hamiltonian is given
\be
H= - c(D) \sum_{j,k} {\ba \phi_j } U_{j,k} \phi_k + h.c. -1/2 \sum_k
|\phi_k|^2.
\ee

The solution of this associated Gaussian model is a crucial step in the
computation of the properties
of the high temperature expansion.

\section{The high temperature expansion}

In the case of the Gaussian model the free energy density can be written
as
\be
\beta F(\beta) = \sum_C W(C) ( \beta  c(D))^{L(C)}/L(C),
\ee
where the sum is done over all the closed lattice circuits with given
starting point; $L(C)$ is the
length of the circuit \cite{PARISI}.

In a model (like the present one) where gauge invariant quantities are
translational  invariant \cite{CA}, we
can chose the origin (and the end) of the circuit at an arbitrary point of
the lattice. In other
cases, like spin glasses,
we must average over all the possible starting points \cite{mpv}.

The previous formula can also be written as
\be
\beta F(\beta)= tr \ln (1+ c(D) \beta \Delta) =\sum_n{ ( \beta
c(D))^n \over n} \N(n) \lan W(C) \ran_n
\ee
where by $\lan W(C) \ran_n$ we denote the average over all the circuits
of length $n$ and by $\N(n)$
the number of (rooted) closed circuits.

Differentiating the previous formulae we obtain a similar result for the
internal energy density:
\be
2 \beta c(D) U(\beta) = \sum_n ( \beta  c(D))^n  \N(n)\lan W(C) \ran_n.
\ee
Here the factor $1/n$ has disappeared.
\subsection{The ferromagnetic case}
\indent
This is the simplest case. We have only to compute $\N(n)$ since $\lan
W(C) \ran_n=1$.

It is evident that $\N(n)=0$ for odd $n$. The first non zero
contributions for small $n$ are
\be
\N(2)= 2D, \ \ \ \ \  \N(4)= 6D (2D-1).
\ee

We could also compute $\N(n)$ using the representation
\be
\N(n)= \int_B d^D p \ (\sum_{\mu=1,D} 2 \cos(p_\mu))^n.
\ee

If we use the correct normalisation of $c(D)$, that gives the
critical temperature  at 1, we
immediately find that when $D \to \infty$
all these contributions vanish. This is a well known fact: in the high
temperature phase in the mean field approximation the internal energy
of a
ferromagnetic system is zero. The
fluctuations contribute only in the subdominant terms of the large $D$
expansion.

This behaviour implies that one should be careful in taking the limit
 $D \to \infty$. Indeed it is easy
to check that in the limit where $n>>D$ one finds
that \cite{parisibook2}

\be
c(D)^n \N(n) \propto n^{-D/2},
\ee
but in the opposite limit $D>>n$ one gets
\be
N(n) \sim (n-1)!!  (2D)^{n/ 2}, \label{fatt}
\ee
and therefore
\be
c(D)^n\N(n) \sim {(n-1)!! \over (2D)^{n/ 2}}.
\ee

The equation (\ref{fatt}) is very simple to understand. In a closed
circuit
for  each step in one direction
there must be a step in the opposite direction. In infinite dimensions all
the steps are taken
in different directions (in a way compatible with this constraint) . The
generic
circuit will be thus
identified by the directions in which these steps are done (we have to
make a choice
$n/2$ times between
these directions) and by the locations of the steps at which two opposite
directions are chosen. In
high dimensions all the steps are done in different directions and in this
way on obtains the
previous formula, i.e. the number of pairing of $n$ objects ($(n-1)!!$),
multiplied by the number of choices for the directions ($(2D)^{n/2}$) .

If we were not aware of the correct normalisation factor and we had
put $c(D)=({1 \over2 D})^{1/2}$ with the
aim of obtaining a non trivial perturbative expansion, we would get the
formula:

\be
\beta U(\beta) = \sum_n (2n-1)!! ( \beta  )^{2n} .
\ee
We would have found in this way that the high temperature expansion
has a zero radius of
convergence. This is not  a surprise\cite{DPS} because in this
scale the critical temperature is at $\beta=0$ and any non zero value of
$\beta$ is already in the
low temperature regime.

In the ferromagnetic case the singularity of the free energy disappears
when $D \to \infty$ in the high temperature expansion with the correct
$c(D)$. This
effect can be easily explained. The ferromagnetic transition is
characterised
by the  building up a singularity at
momentum $k=0$ in the two point correlation function.
The free energy in the high temperature phase is given by
\be
 f(\beta) \propto \int_B d^Dk \ln(1-\beta \sum_{\nu=1,D}
\cos(k_\nu)/(2D)),
\ee
where the integral is done over the first Brillouin zone.

When $D \to \infty$ the region of momenta
near the origin has a vanishing weight and its contribution to the
singularity disappears.  We can see a transition in the  specific heat in
the
limit of infinite dimensions only if the directions  of the most relevant
modes  are not orthogonal to the
boundary of the Brillouin zone, where the measure is concentrated in
momentum space.

\subsection{Spin Glasses}

In this case we will compute the spectrum of the random Laplacian.
This
can be done in the
infinite dimensions limit since we recover the old problem of computing
the
spectrum of a random matrix,
which is given by a semicircle law\footnote
{One could also use the replica approach.}.
Instead of
using directly this result we  prefer to follow a
diagrammatic approach.

In this case the $U$'s have zero average and are random elements of the
$U(1)$ group. After the
average over all the possible starting points, $W(C)$ gets  contributions
only from those circuits for
which for any step going from $i$ to $k$ there is a step going from $k$
to $i$. In other words we
must sum only over {\sl backtracking} circuits.

Let us count the number of these circuits in infinite dimensions. We
must to compute
\be
G_{2n} = \lim_{D\to\infty}(2D)^{-n} \N(2n)  \lan W \ran _{2n}.
\ee

 It is easy to check that for $n=1$ we do not get any new contribution
with respect to the previous
case  and $G_1=1$.

For larger values of $n$ a more detailed computation must be done. At
this end
it is convenient to denote by
$a,\ b,\ c..$ one of the different $2D$ possible directions in which a step
could be done.

In the case $n=2$ we have 3!! circuits which differs for the ordering
possibilities:
\be
a a b  b  \ ab ba  \ a bab,
\ee
where it is implicit that the second identical letter denotes a back step in
the opposite direction
of the first one. We do not attach any meaning to the letters $a$ or $b$:
we could have written $a a b  b $ or $bbaa$ indifferently. In both case
the second step and the fourth steps are in the opposite direction of the
first and of the third step respectively.

Each of the  3!! choices correspond to $(2D)^2$ lattice
circuits (we neglect subleading terms for large $D$).
The first two are
backtracking circuits the second is not. We thus find $G_2=2$.

In the case $n=3$ we have 5!! circuits which differs for the ordering
possibilities. We list here
all the backtracking ones:
\be
aabbcc \ abbcca \ abccba \ aabccb \ abccab.
\ee
Therefore $G_3=5$.
It is easy to verify that a circuit is backtracking if and only if the
corresponding word may be
reduced to the null one by subsequent elimination of consecutive
identical letters.

The computation of $G_n$ can be thus cast under the following
graphical form. For each given word, we put its $2n$ letters (two by
two equal), on
a circle  starting from a given point, in the same
order of the
letters of the corresponding word. We connect those points which
have identical letter by a
line and we count the number of intersections of the lines. This
number is topological invariant
and it does not depend on the point where the letter have been put on the
circle ,
but only on their order.

 We can associate to each word the number of intersections. Let us call
$I_n(m)$ the number of
words which have $m$ intersections ($m\le n(n-1)/2$). It is easy to
check that
\be
I_n(0)=G_n.
 \ee
Indeed only in the case in which the resulting diagram is planar, the
diagram may be reduced to zero
by removing consecutively equal letters.

The combinatorial problem of computing $I_n(0)$ has been solved
\cite{BIPZ} in the past\footnote{
The result is a by-product of the formula relating the generating
functionals of the connected and of
the disconnected functions. }.
After a short computation one finds:
\be
I_n(0)= 4^n{\Gamma(n+1/2) \over \Gamma(1/2) \Gamma(n+2)}
\ee

The result of the computation can also be written is a slightly different
form. We  consider an Hilbert space, and a base
($|m\ran$ on this Hilbert  space, where  $m$ ranges in the interval $[0
-\infty]$). We define on this
space  two  shift operators $\R$ and $\LL$:
\bea
\R |m \ran = |m +1\ran \\
\LL|m \ran = |m -1\ran,
\eea
where $|-1\ran$ is identified with the null vector.

These two operators satisfy the relation,
\be
\LL\R =1,
\ee
which is a particular case (for $q=0$) of the $q$-deformed
commutation relations
\footnote
{In the case $q=1$ we have Bosonic commutation relations, for $q=-1$
we have Fermionic commutation
relations and for $q=\exp(i\theta)$ anionic commutation relations. Some
applications of the anionic commutation relations can be found in
\cite{VI,LS} and references therein.}: \be \LL \R - q \R \LL=1.
\ee

It is easy to see that
\be
G_n = \lan 0| (\R +\LL)^{2n}|0\ran,
\ee
where the state $|0\ran$ could also be characterised the condition
\be
\LL|0\ran =0.
\ee

The existence of these two other formulations should not be a surprise.
The condition of zero
intersection implies that the diagram is planar and the theory of random
matrices may be reformulated
in terms of planar diagrams.  The theory of random matrices can also
be formulated in terms of the
orthogonal polynomials respect to a given measure \cite{Bessis}
and in this contest it is well known that the shift operators play a crucial
role \cite{PA79}.

We finally find that
\be
1+\beta U(\beta) =\sum_n (4\beta^2)^n{\Gamma(n+1/2) \over \Gamma(1/2)
\Gamma(n+2)} = {2\over \pi}\int_{-2}^{2} d\la
{(1-\la^2/4)^{1/2} \over (1+\beta\la)}\label{SG}
  \ee

There is a transition at $\beta=1/2$, which is characterised  a singularity
of the specific heat of
the form $(\beta_c-\beta)^{-1/2}$. In other words the critical exponent
$\al$ is equal to $1/2$.

Equation (\ref{SG}) gives the result for spin glasses in the Gaussian
approximation.
Starting from it one can obtain
the more familiar results for the Ising spin glass or for  the spherical
spin
glass.

\subsection {Josephson junctions in Magnetic Field}

In this case we need  at first to compute the function
\be
G_{n}(B) = \lim_{D\to\infty}(2D)^{-n} \N(2n)  \lan W \ran_n.
\ee
We will follow the strategy of first dividing the circuits into classes
corresponding to different
words of $2n$ letters (as in the previous case) and to evaluate the
contribution of each class.

Let us start by computing $G_{2}(B)$ (it is trivial that $G_{1}(B)=1$).
The
backtracking circuits which correspond to
the {\sl planar} diagrams, (the corresponding words are  $aabb$ and
$abba$) give a contribution 1 each.
More generally we can define the area of a circuit as the minimal area
of a surface of lattice
plaquettes which have that circuit as boundary. Backtracking circuits
can be characterised as area
zero circuits.

For large $D$ the
word $abab$ corresponds to $(2D)^2$ circuits with area 1. For
half of them the signed area
(defined in eq. \ref{SIGNED}) $S(C) $ is equal to 1, for the other half
is
equal to -1. If we recall that
$W(C)=\exp(i\Phi(C))$, the contribution of these circuits average to
$\cos(B)$. We finally find
\be
G_{4}(B)= 2+q,
\ee
where
\be
q= \cos (B).
\ee

Generally speaking  each different word of length $2n$ is associated to
$(2D)^n$ circuits having  the same area. The signed
area of these circuits having the same area ($A$) is different. In a large
number of
dimensions (in  the generic case where all the independent steps are
done
in different directions) the projected signed areas $S_{\mu,\nu}$ take
only the
values 0 or $\pm 1 $ and  \be
\sum |S_{\mu,\nu}|= A.
\ee

If we average over all the possible orientations of the lattice the
contribution coming from the
circuits having the same word, we find that the average value of $\lan
W(C)\ran$ depends only on $A$ and it is given by
\be
\lan W(C)\ran _A = ({\exp(iB)+\exp(-iB) \over 2})^A= q^A.
\ee

We finally find that
\be
G_n(B) = \sum_w q^{A(w)},
\ee
where the sum is taken over all words of $2n$ letters and $A(w)$ is the
area associated to each of
these words.

We now show  that the area of a of the circuit is exactly equal to the
number  of intersections of
the lines connecting equal letters in the corresponding diagram. We can
decrease the area by an unity by
interchanging two letters. For
example
\be
A(acdefbacdefb)=A(acdefabcdefb)+1.
\ee
Indeed the area of the projection on the $a- b $ plane goes from 1 to 0
and the projected area
on the other planes is the same in the two circuits corresponding to the two
words. The same braiding
operation decreases the number of intersections by 1. By subsequent
operations of the previous kind we
can arrive to the zero intersections case (planar diagrams) by decreasing
each time both the area
and the projection by an unity. We have already remarked the
relation between the number of
planar diagrams and the coefficient of the high temperature expansion
for spin glasses ($G_n=G_n(0)$).

We have thus transformed the problem of computing the high
temperature expansion into a combinatorial
problem, although not very easy, which generalise the computation of
planar diagrams.  The solution of
this problem will be presented in the next section.

\section{The $q$-deformed harmonic oscillator  plays a role}

We have reduced the problem of evaluating the high temperature
expansion for the Gaussian model
in presence of a magnetic field to the computation of the number of
words of $2n$ letters, two by
two equal, such that the number of intersections in the corresponding
diagram is equal to a given
number.

We claim that
\be
G_n(B) = \sum_w q^{A(w)}= \lan 0| X^{2n}|0\ran,\label{MAGIC}
\ee
where
\be
X= \R_q +\LL_q,
\ee
and the operators $\LL$ and $\R$  satisfy the commutation relation of a
$q$-deformed harmonic
oscillator:
\be
\LL_q \R_q - q \R_q \LL_q=1.
\ee
Therefore $\LL_q$ may be identified with the distruction operator and
$R_q$ with the creation operator
for a $q$-deformed harmonic oscillator. For $q=1$ we recover the
ferromagnetic case, for $q=-1$ the
fully frustrated case and for $q=0$ the spin glass case.

These operators may be represented as:
\bea
\R_q |m \ran =[m]_q^{1/2} |m +1\ran, \\
\LL_q|m \ran = [m-1]_q^{1/2} |m -1\ran,
\eea
where
\be
[m]_q=(1-q^{m+1})/(1-q),
\ee
and   $m$ ranges in the interval $[0 -\infty]$.
In the limit $q \to 1$ we obtain the usual Bosonic oscillator and we
recover the usual formulae.

It is a simple matter of computation to verify that eq.(\ref{MAGIC})
gives
\bea
G_1(B)=1,\ G_1(B)=2+q, \ G_3(B)=5+6q+3q^2+q^3, \\
 G_4(B)= 14 + 28q + 28q^2 + 20q^3 + 10q^4 + 4q^5 + q^6.
\eea
These results coincide with the output of an explicit enumeration of the
diagrams.

We have not been able of finding a neat proof of eq.(\ref{MAGIC}).
However we have checked its
validity in many special cases (large $q$, small $q$, $q=1$, $q=0$ and
$q=-1$) and we are convinced
of its validity.

 Intuitively
eq.(\ref{MAGIC}) tells us that when we use the {\sl Wick} theorem
for
$q$-deformed harmonic
oscillators, we must bring together the different terms we
contract and for each term we get
a factor $q$ to the power of the number of object we have to cross.

If we use this result, we finally find the quite simple formula:
\be
1+ \beta U(\beta)=\lan 0| {1 \over 1+\beta X} |0\ran_q,
 \ee
which gives a remarkable connection among the high temperature
behaviour of the Gaussian model and
the $q$ deformed harmonic oscillator.

In this way we have reduced the combinatorial problem of computing
the high temperature expansion to
an algebraic problem.

\section{Near the critical transition}

The problem now is reduced to the computation of the spectrum of the
operator $X$ of the
$q$-deformed harmonic oscillator. The computation is apparently non
trivial. We are however
interested to the computation of the spectral density near the largest
eigenvalues.

A simple case is $q=1$, where the operator $X_q$ is not bounded and
the high temperature expansion is
divergent. In this case $X$ has a continuum spectrum and the highest
eigenvalues of $X$ are
concentrated in the large $m$ region. Let us assume that this feature is
valid for $q$ inside the
interval $[-1,1]$. One finds that
\be
\LL_q \sim (1-q)^{-1/2}\LL, \ \R \sim (1-q)^{-1/2}\R.
\ee
when the operator is applied to a state $|m\ran$ in the region of large
$m$. ($\LL$ and $\R$ are the two shift operators for $q=0$ which are
used in
the planar case).

The difference among $\LL_q$ and  $(1-q)^{-1/2}\LL$ can be seen
only when the two operators act
on a state of low $m$.  It is very reasonable to assume that the spectral
radius and
the spectral density near the
maximum eigenvalues is the same in the two case. We have verified
numerically
that this conjecture is
consistent (at least for $q$ not too close to 1) by estimating the spectral
density of $X_q$ in
subspaces of various size ($m< M$, with $M$ up to a 300).

We find therefore that the critical temperature is given by
\be
\beta_c={(1-q)^{1/2}\over 2},
\ee
which is the inverse of the spectral value of $X$, i.e.
\be
|X|^2={4 \over   (1-q)}.
\ee
The behaviour of the spectral density near the edge is the same as for
the random matrix
model, i.e. in spin glass. In this way we find the same critical exponents
as in spin glasses in the
Gaussian approximation.

A possible physical interpretation is the following. In computing the
internal energy one has to
sum over all the closed circuits. Circuits with large physical area
average to zero
and only fattened backtracking circuits survive. The situation is very
similar to spin glasses,
where only backtracking circuits contribute, the only effect being a
renormalization of the
temperature\footnote
{The previous results imply that when $n$ and $m$ goes both to
infinity at fixed ratio one finds $I_n(m)=I_n(0){ (n+m)! \over n! m!}
f(m/n)$. It is quite possible that this simple result has a direct proof.}

\section{The issue of exchanging limits}
A very serious problems in assessing the relevance of these results is
related to
the exchange of the limits $D\to \infty$ and $\beta \to \beta_c$. If we
exchange
the limits  we become blind to any singularity whose strength vanishes
in the
limit $D \to \infty$.  Sometimes this exchange is quite
justified, sometimes it leads to disaster \cite{PZ},\cite{DPTV}.

The cases $q=1$ and $q=-1$ are particularly instructive. The case
$q=1$ has been already discussed.
The case $q=-1$  is quite interesting. We notice the following facts.
\begin{itemize}
\item
The spectrum of the lattice Laplacian for the fully frustated model is well
known \cite{DPTV}. A simple way to compute it consists in using the
relation among the Gaussian fully  frustrated
model and the naive Wilson
Fermions on the lattice \cite{KS}. Indeed let us start from the
Hamiltonian of the
naive Wilson Fermions
\be
H=\sum_i (\sum_{\mu} (\beta( \ba \psi(i+\hat \mu)-\ba \psi(i-\hat \mu))
\gamma_\mu \psi(i)) +\ba \psi(i)
\psi(i)), \ee
where $\hat \mu$ is the versor in $\mu$ direction, the $\gamma_\mu$
are the appropriate
Dirac gamma matrices in $D$ dimensions (which satisfies the usual
algebra) and the $\psi$ are the
spinors on which these matrices act. For even $D$ the
gamma matrices may be taken to have
dimension $2^D/2$. In order to simplify the notation we have not
indicated the
spinorial indices.
 If we introduce the field
\be
\phi(i)=\prod_{\mu=1,D} \gamma_\mu^{i_\mu} \psi(i),
\ee
it is well known fact that the lattice Dirac operator reduces to the
Laplacian of a fully frustrated
model.
\item
The previous remark implies that for $q=-1$ one has in the Gaussian
approximation (with the
appropriate rescaling of $\beta$):
\be
1+ \beta U(\beta)= \int_B d^Dk {1 \over (1-2\beta^2 \sum_{\nu=1,D}
\sin^2(k_\nu)/D)},
\ee
for all even values of the dimensions.
\item
If we send $D$ to infinity we find that
\be
1+ \beta U(\beta)={1\over 1-\beta^2},
\ee
in perfect agreement with the direct computation. (In this case the
creation and annihilation operators
act on a two dimensional Fermionic space.)
\item
In any finite dimensions  \cite{DPTV} the closest singularity to the origin of
the function $U(\beta)$ is  located at $\beta^2=1/2$, which
corresponds to the integration point where all the momenta are at the
boundary of the Brillouin zone
(i.e. ($k_\mu)= \pm \pi/2$).
\item
 In infinite dimensions the function $\beta_c(q)$ is discontinuous at $q=-
1$. Indeed
\be
\lim_{q\to -1} \beta^2_c(q) =1/2 \ne \beta^2_c(-1) =1
\ee
 \end{itemize}
As already found in \cite{DPTV}, at $q=- 1$ the limit $D \to \infty$ of
$\beta_c$ is smaller by a factor 2 of the value of $\beta_c$ obtained from the
high temperature expansion computed directly ad $D=\infty$.  However this
difficulty seems to be confined at $q=-1$. If we first take the limit
$D\to\infty$ at  $q\ne 1$, we recover the correct critical point for the $q
=1$ case.

In other words, if we first compute the critical temperature at
$D=\infty$
for $q\NE - 1$, we obtain the correct value of the critical temperature
at
$q=-1$, while we would get the wrong results if we perform the limit
$D\to\infty$ directly at $q=-1$. By consistency we find that  the
prefactor in
front of the nearest discontinuity vanishes when $q \to -1$ so that for
$q=-1$
this  singularity disappears.

It seems that we are free to conjecture that (apart from two well
understood
problems at $q=-1$ \cite{DPTV} and
$q=1$ \cite{DPS}) the correct value of the critical temperature is obtained
when we  send firstly $D$ to infinity.
A numerical verification of the validity of this conjecture may be
attempted for $q=0$ or $\pm{1 \over
2}$, where $\beta_c$ can be computed by diagonalizing matrices of size
$2^D$ or $3^D$ respectively.
 \section{Open problems}

Let us  suppose that the  difficulties discussed in the previous section are
not serious. We still
face the problem of presenting a full computation of the high
temperature expansion in the $XY$ model.
We must include high order terms which come from the fact that
the distribution of the spins
is not Gaussian. In the case of spin glasses these corrections are relevant;
however they are
identical in the Ising, $XY$ and spherical model. In this last case they
can be computed by tuning
the coefficient of the quadratic term in such way that the spherical
constraint is satisfied.

We have not checked that this happens  also in our case, but it seems
rather plausible. If this
argument is correct, the knowledge of the Gaussian propagator is
sufficient to reconstruct the high
temperature expansion.

What happens in a finite number of dimensions is not clear. The first
step is to verify
if the equality of the
two model survives in perturbation theory. Also if this check is satisfied
one should be very careful
because of non  perturbative effects. It seems to me rather likely, but I
do not have solid arguments in
this direction, that for rational $B$ the critical theory should behave
differently from spin
glasses, and the only hope for having a spin glass like behaviour is for
generic irrational $B$. It would be vry interesting to connect this approach
with the results obtained in two dimensions, where quantum groups have been
used to compute the spectrum\cite {WZ}.

The possibility of having a spin glass behaviour
for this non random  system \cite{MPR} is  fascinating and deserves
more careful investigations.

\section{Acknowledgements}
It is a pleasure for me to thank V. Anders  for suggesting to me this
field of
investigation and for a very
useful discussion. I am also happy to thank for useful suggestions G.
Immirzi, E. Marinari, S. Sciuto and G.
Vitiello.

\section{Appendix}
In this short appendix I report on some numerical
findings that I have obtained on the behaviour of the spectral radius of
$X$ as function of $\theta$
for $q=\exp(i 2 \pi\theta)$. In this case I find a function which is
discontinuous at all the
rational points, but the discontinuity vanishes when the rational point
becomes irrational.

	If we apply the previous formulae we find that the spectral radius
of $X^2$ should be
\be
{4 \over |1-q|}= \left( {4\over sin^2( \pi \theta)} \right) ^{1/2}
\ee

The argument breaks down for rational $\theta$. Indeed if $\theta=r/s$,
with both $r$ and $s$ integer
($r$ and $s$ are the smallest integer for which have this property) $X$
reduces to a finite
dimensional operator of size $s$. In this case the previous formula is
not correct. However
in the limit where $s$ goes to infinity it seems to become correct again.
This can be seen by
considering the function $R(\theta)$, defined as

\be
|X| (\theta)^4= {4\over sin^2( \pi \theta)} (1-{ \pi^2 \over 2 s^2
})+R(\theta)
\ee

The function $R(\theta)$ is the difference among the analytic
continuation
of the value of the  spectral radius from $|q|$ less than 1
and the actual spectral radius (apart the  presence of a
multiplicative  factor which goes to zero as
$s^{-2}$ when $s \to \infty$ at fixed $\theta$).

I have computed the function $R(\theta)$ for all rational with $s\le
21$ ($70$ cases) and I have found that goes to zero fast with $s$ (quite
likely as $s^{-2}$).  It seems likely that the function $R(\theta)$ is
discontinuous at rational points, but the value of the discontinuity
goes to zero when the rational becomes irrational (i.e. when $s \to
\infty)$.

Unfortunately I am not aware of a physical interesting model in which
the properties of $X$ for complex $q$ enter. This appendix should be
considered as a curiosity.

\end{document}